\documentclass[%
notitlepage,nofootinbib,superscriptaddress,
showpacs,preprintnumbers,
amsmath,amssymb,
prd,onecolumn%
]{revtex4-1}

\usepackage{epsfig,subfigure}
\usepackage{graphicx}
\usepackage{bbm}
\usepackage[colorlinks=true,citecolor=blue]{hyperref}
\draft 
\begin{document}
\author{Radha Raman Gautam}%
 \email{gautamrrg@gmail.com}
  \affiliation{Department of Physics, Himachal Pradesh University, Shimla 171005, INDIA.}%
\author{Madan Singh}
 \email{singhmadan179@gmail.com}
  \affiliation{Department of Physics, Panjab University, Chandigarh 160014, INDIA.}%
\author{Manmohan Gupta}%
 \email{mmgupta@pu.ac.in}
  \affiliation{Department of Physics, Panjab University, Chandigarh 160014, INDIA.}
  
\title{Neutrino mass matrices with one texture zero and a vanishing neutrino mass}

\begin{abstract}
Assuming Majorana nature of neutrinos, we investigate the singular one texture zero neutrino mass matrices in the flavor basis. We find that for the normal mass ordering with $m_1=0$, all the six one texture zero classes are now ruled out at 3$\sigma$ confidence level, whereas for inverted mass ordering with $m_3=0$ only four classes out of total six can accommodate the latest neutrino oscillation data at 3$\sigma$ confidence level. Moreover, only two classes can accommodate the present data at 1$\sigma$ confidence level. We examine the phenomenological implications of the allowed classes for the effective Majorana mass, Dirac and Majorana CP-violating phases. Working within the framework of type-I seesaw mechanism, we present simple discrete Abelian symmetry models leading to all the phenomenologically allowed classes.
\end{abstract}
\maketitle
\section{Introduction}
In the last three years, T2K, MINOS, Double Chooz, Daya Bay and RENO experiments \cite{1,2,3,4,5} have established a non-zero and relatively large value of the reactor mixing angle $\theta_{13}$. The result has further motivated experimentalists to pin down the long-standing problem of CP violation and mass ordering in neutrino sector. Recent global fits of neutrino oscillations \cite{6,7}, have reported best fit points and 1$\sigma$ errors on the Dirac CP-violating phase $\delta$. 
On the theoretical side, in order to explain the pattern of neutrino masses and mixing, several ideas have been proposed in the literature which reduce the number of free parameters in neutrino mass matrix, e.g. some elements of the neutrino mass matrix are considered to be zero \cite{8,9,10,11} or equal \cite{12} or both the possibilities are taken together \cite{13}. Similarly, some co-factors of the neutrino mass matrix are considered to be zero \cite{14} or equal \cite{12} or both \cite{15}. The analysis of two texture zero neutrino mass matrices in the flavor basis restricts the number of experimentally compatible classes to seven. The phenomenological implications of one texture zero neutrino mass matrices have also been investigated in the literature \cite{16,17,18} and it has been found that all the six possible classes with one texture zero in the neutrino mass matrix are experimentally viable.\\ 
In the flavor basis where the charged lepton mass matrix is diagonal, the Majorana neutrino mass matrix, being complex symmetric, contains six independent entries. If one of the elements is assumed to be zero, then we have six possible one texture zero classes which are shown in Table \ref{tab1}.
\begin{table}[htp]
\begin{center}
\begin{tabular}{|c|c|c|}
\hline $P_1$ & $P_2$ & $P_3$  \\ 
 \hline $\left(
\begin{array}{ccc}
0 & \times & \times \\
\times & \times & \times\\
\times& \times & \times \\
\end{array}
\right)$ & $\left(
\begin{array}{ccc}
\times& \times & \times \\
\times & 0 & \times\\
\times& \times & \times \\
\end{array}
\right)$ & $\left(
\begin{array}{ccc}
    \times& \times & \times \\
  \times & \times & \times\\
  \times& \times & 0 \\
\end{array}
\right)$  \\ 
\hline $P_4$ & $P_5$ & $P_6$  \\ 
\hline  $\left(
\begin{array}{ccc}
    \times& 0 & \times \\
  0 & \times & \times\\
  \times& \times & \times \\
\end{array}
\right)$&$\left(
\begin{array}{ccc}
    \times& \times & 0 \\
  \times & \times & \times\\
  0& \times & \times \\
\end{array}
\right)$ &  $\left(
\begin{array}{ccc}
   \times& \times & \times \\
  \times & \times & 0\\
  \times& 0 & \times \\
\end{array}
\right)$ \\ \hline
\end{tabular}
\caption{\label{tab1} Possible structures of neutrino mass matrices having one texture zero. $\times$ denotes the non-zero elements.}
\end{center}
\end{table}
The condition of a single texture zero in the neutrino mass matrix is less restrictive (and hence less predictive) than the condition of two texture zeros. However, one can further reduce the number of free parameters of one texture zero Majorana neutrino mass matrices by considering one of the neutrino masses to be zero, which is still an experimentally viable scenario. In Refs. \cite{17,19}, particular attention has been paid to neutrino mass matrices with one texture zero and a vanishing neutrino mass, which is termed as singular one texture zero model. Recently, individual classes of one texture zero with a vanishing neutrino mass have been studied by some authors, e.g. in Ref. \cite{20} the authors have considered classes $P_4$ and $P_5$ with the additional constraint of a new kind of constrained sequential dominance (CSD2) \cite{20, 21}. In Ref. \cite{22} classes $P_1$, $P_4$ and $P_5$ have been explored with the motivation of linking the ratio of mass-squared  differences and the (1,3) element of the neutrino mixing matrix. Class $P_4$ has also been studied in Ref. \cite{23} and class $P_3$ has been studied in Ref. \cite{24}. In Ref. \cite{25} all the singular one texture zero classes along with other new texture structures have been obtained by systematically scanning the zeros of the Dirac and the right-handed Majorana neutrino mass matrices within the context of type-I seesaw mechanism \cite{26}.\\
Classes $P_4$, $P_5$ and $P_6$ have also been derived in Ref. \cite{27} in the minimal type-I seesaw model considering only two right-handed heavy Majorana neutrinos. Further, these minimal models have been reinvestigated in Ref. \cite{28} by keeping minimum number of parameters which can lead to successful leptogenesis and considering the relatively large value of $\theta_{13}$. Recently, the renormalization group effects on neutrino mixing parameters corresponding to classes $P_4$, $P_5$ and $P_6$ have been considered in Ref. \cite{29}.\\
In view of the refined measurement of reactor mixing angle $\theta_{13}$ and with the motive of carrying out a complete phenomenological analysis of all the classes of singular one texture zero, we investigate neutrino mass matrices with one texture zero and a vanishing neutrino mass. Working within the framework of type-I seesaw mechanism \cite{26}, we construct simple neutrino mass models based on $Z_8$ discrete symmetry which lead to the experimentally allowed classes studied in the present work.\\
The rest of the paper is structured as follows: In section 2, we discuss the methodology employed to obtain the constraint equations for one texture zero. Section 3 is devoted to numerical analysis. In section 4, we give the details of symmetry realization of all the allowed classes. In section 4, we summarize our work.
\section{Methodology}
The effective Majorana neutrino mass matrix $(M_{\nu})$ contains nine parameters which include three neutrino masses ($m_{1}$, $m_{2}$, $m_{3}$), three mixing angles ($\theta_{12}$, $\theta_{23}$, $\theta_{13}$) and three CP-violating phases ($\delta$, $\rho$, $\sigma$). In the flavor basis, the Majorana neutrino mass matrix can be expressed as,
\begin{equation}
 M_{\nu}=VM^{\textrm{diag}}V^{T}
\end{equation}
where $M^{\textrm{diag}}$ = diag($m_{1}$, $m_{2}$, $m_{3}$) is the diagonal matrix of neutrino masses and $V$ is the flavor mixing matrix. The above equation can be re-written as
\begin{equation}
M_{\nu}=U\left(
\begin{array}{ccc}
    \lambda_{1}& 0& 0 \\
  0 & \lambda_{2} & 0\\
  0& 0& \lambda_{3} \\
\end{array}
\right)U^{T}.
\end{equation}
where
$\lambda_{1} = m_{1} e^{2i\rho},\lambda_{2} = m_{2} e^{2i\sigma} ,\lambda_{3} = m_{3}.$
For our analysis, we consider the following parametrization of $V$ \cite{30}:
\begin{equation}
V \equiv UP=\left(
\begin{array}{ccc}
 c_{12}c_{13}& s_{12}c_{13}& s_{13} \\
-c_{12}s_{23}s_{13}-s_{12}c_{23}e^{-i\delta} & -s_{12}s_{23}s_{13}+c_{12}c_{23}e^{-i\delta} & s_{23}c_{13}\\
 -c_{12}c_{23}s_{13}+s_{12}s_{23}e^{-i\delta}& -s_{12}c_{23}s_{13}-c_{12}s_{23}e^{-i\delta}& c_{23}c_{13} \\
\end{array}
 \right)P,
\end{equation}
where, $c_{ij} = \cos \theta_{ij}$, $s_{ij}= \sin \theta_{ij}$. Here, $U$ is a 3 $\times$ 3 unitary matrix consisting of three flavor mixing angles ($\theta_{12}$, $\theta_{23}$, $\theta_{13}$) and one Dirac CP-violating phase $\delta$ and $P$ = diag($e^{2i\rho}, e^{2i\sigma}, 1$), is a diagonal phase matrix consisting of two Majorana CP-violating phases $\rho$ and $\sigma$.\\
If one of the elements of $M_{\nu}$ is considered zero, \textit{i.e.} $M_{lm} = 0$, we obtain the following constraint equation
\begin{equation}
\sum_{i=1,2,3}U_{li}U_{mi}\lambda_{i}=0
\end{equation}
where $l$, $m$ run over e, $\mu$ and $\tau$. Two independent mass-squared differences $\delta m^{2}$ (solar) and $\Delta m^{2}$ (atmospheric) are defined as
\begin{equation}
 \delta m^{2}=(m_{2}^{2}-m_{1}^{2}),\;
 \end{equation}
 \begin{equation}
  \Delta m^{2}=|m_{3}^{2}-m_{2}^{2}|.
 \end{equation}
The ratio of above mass-squared differences is given by
\begin{equation}
 R_{\nu}=\frac{\delta m^{2}} {|\Delta m^{2}|} \ .
\end{equation}
If we consider one of the neutrino masses to be zero then since, $m_{2}>m_{1}$ has already been confirmed by solar neutrino oscillation data \cite{31,32}, $m_{2}$ cannot be equal to zero. Thus, we are left with two possibilities where either $m_{1}$ or $m_{3}$ can vanish corresponding to normal ($m_1 = 0, m_2 < m_3$) or inverted ($m_1 < m_2, m_3 = 0$) mass ordering, respectively. The vanishing lowest neutrino mass along with one texture zero condition put constraints on the parameter space of neutrino masses, neutrino mixing angles and CP violating phases.\\
In case of one texture zero neutrino mass matrices, there exists a permutation symmetry between certain classes. This corresponds to permutation of the 2-3 rows and 2-3 columns of $M_{\nu}$. The corresponding permutation matrix is
\begin{center}
\begin{equation}
 P_{23} = \left(
\begin{array}{ccc}
    1& 0& 0 \\
  0 & 0 & 1\\
  0& 1& 0 \\
\end{array}
\right).
\end{equation}
\end{center}
As a result of permutation symmetry between different classes, one obtains the following relations among the oscillation parameters
\begin{center}
\begin{equation}
\theta_{12}^{X}=\theta_{12}^{Y}, \ \ 
\theta_{23}^{X}=90^{\circ}-\theta_{23}^{Y},\ \ 
\theta_{13}^{X}=\theta_{13}^{Y}, \ \ \delta^{X}=\delta^{Y} -  180^{\circ},
\end{equation}
\end{center}
where X and Y denote the classes related by 2-3 permutation. The following one texture zero classes are related via permutation
symmetry
\begin{equation}
P_{2} \leftrightarrow P_{3}, \ \ \  P_{4} \leftrightarrow P_{5} \ .
\end{equation}
Classes $P_1$ and $P_6$ transform unto themselves under the action of $P_{23}$.
\begin{center}
\textbf{Case I: $m_{1}=0$ (normal mass ordering)}
\end{center}
Using Eq. (4), we get the following expressions for the neutrino mass ratio $\left(\frac{m_{2}}{m_3}\right)$ and the Majorana phase $\sigma$
\begin{equation}
\frac{m_{2}}{m_{3}}= \frac{|U_{l3} U_{m3}|}{|U_{l2} U_{m2}|},
\end{equation}
\begin{equation}
\sigma=\frac{1} {2} \textrm{arg}\bigg(-\frac{U_{l3} U_{m3}}{U_{l2} U_{m2}}\bigg).
\end{equation}
Since $m_{1}$ is zero, therefore, Majorana phase $\rho$ becomes unphysical in this case. Using Eqs. (5) and (6), neutrino masses ($m_{1}$, $m_{2}$, $m_{3}$) can be expressed in terms of experimentally known mass-squared differences ($\delta m^{2}$, $\Delta m^{2}$) as
\begin{equation}
m_{1}=0, \ \  m_{2}=\sqrt{\delta m^{2}}, \ \  m_{3}=\sqrt{\delta m^{2}+\Delta m^{2}}.
\end{equation}
Hence, we obtain
\begin{equation}
 \frac{m_{2}} {m_{3}}=\sqrt\frac{R_{\nu}}{1+R_{\nu}}.
\end{equation}
Using Eqs. (11) and (14), we can express $R_{\nu}$ in terms of mixing angles ($\theta_{12}$, $\theta_{23}$, $\theta_{13}$) and
Dirac CP-violating phase ($\delta$) as 
\begin{equation}
 R_{\nu}=\bigg(\frac{|U_{l2}U_{m2}|^{2}} {|U_{l3}
 U_{m3}|^{2}} -1\bigg)^{-1}.
\end{equation}
\begin{center}
\textbf{Case II: $m_{3}=0$ (inverted mass ordering)}
\end{center}
 The expressions for neutrino mass ratio $\left(\frac{m_{2}}{m_{1}}\right)$ and Majorana phase difference $(\rho$ - $\sigma)$ are given by
\begin{equation}
\frac{m_{2}}{m_{1}}=  \frac{|U_{l1} U_{m1}|}{|U_{l2} U_{m2}|},
\end{equation}
\begin{equation}
\rho-\sigma =\frac{1} {2}  arg\bigg(-\frac{U_{l2} U_{m2}}{U_{l1} U_{m1}}\bigg).
\end{equation}
In the case of Inverted mass ordering, the phase difference $(\rho-\sigma)$ is the relevant physical phase. From Eq. (17), it is
clear that Majorana phases ($\rho,\sigma$) are linearly co-related. The neutrino mass spectrum for inverted mass ordering is given by
\begin{equation}
m_{1}=\sqrt{\Delta m^{2}-\delta m^{2}}, \ \  m_{2}=\sqrt{\Delta m^{2}}, \ \  m_{3}=0. 
\end{equation}
The mass ratio $\left(\frac{m_{2}}{m_{1}}\right)$ is related to $R_{\nu}$ and is given by
\begin{equation}
 \frac{m_{2}} {m_{1}}=\frac{1} {\sqrt{1-R_{\nu}}}.
\end{equation}
Using Eqs. (16) and (19), we can express $R_{\nu}$ in terms of mixing angles ($\theta_{12}$, $\theta_{23}$, $\theta_{13}$) and
Dirac CP-violating phase ($\delta$) as 
\begin{equation} 
R_{\nu}=1- \bigg|\frac{U_{l2}U_{m2}} {U_{l1} U_{m1}}\bigg|^{2} .
\end{equation}
The expression for Jarlskog rephasing parameter $J_{CP}$, which is a measure of CP violation, is given by
\begin{equation}
J_{CP}=s_{12} c_{12} s_{23}  c_{23} s_{13}  c_{13}^{2} \sin\delta.
\end{equation}

\begin{table}
\begin{small}
\begin{center}
\begin{tabular}{|c|c|c|c|c|}
  \hline
  Parameter& Best Fit & 1$\sigma$ & 2$\sigma$ & 3$\sigma$ \\
  \hline
   $\delta m^{2}$ $[10^{-5}eV^{2}]$ & $7.60$& $7.42$ - $7.79$ & $7.26$ - $7.99$ & $7.11$ - $8.18$ \\
   \hline
   $|\Delta m^{2}_{31}|$ $[10^{-3}eV^{2}]$ (NO) & $2.48$ & $2.41$ - $2.53$ & $2.35$ - $2.59$ & $2.30$ - $2.65$\\
   \hline
  $|\Delta m^{2}_{31}|$ $[10^{-3}eV^{2}]$ (IO) & $2.38$ & $2.32$ - $2.43$ & $2.26$ - $2.48$ & $2.20$ - $2.54$ \\
  \hline
  $\theta_{12}$ & $34.6^{\circ}$ & $33.6^{\circ}$ - $35.6^{\circ}$ & $32.7^{\circ}$ - $36.7^{\circ}$ & $31.8^{\circ}$ - $37.8^{\circ}$\\
  \hline
  $ \theta_{23}$ (NO) & $48.9^{\circ}$ &$41.7^{\circ}$ - $50.7^{\circ}$  & $40.0^{\circ}$ - $52.1^{\circ}$ & $38.8^{\circ}$ - $53.3^{\circ}$ \\
  \hline
  $\theta_{23}$ (IO)& $49.2^{\circ}$ & $46.9^{\circ}$ - $50.7^{\circ}$ & $41.3^{\circ}$ - $52.0^{\circ} $& $39.4^{\circ}$ - $53.1^{\circ}$ \\
  \hline
  $\theta_{13}$ (NO) & $8.6^{\circ}$ & $8.4^{\circ}$ - $8.9^{\circ}$ & $8.2^{\circ}$ - $9.1^{\circ}$& $7.9^{\circ}$ - $9.3^{\circ}$ \\
  \hline
  $\theta_{13}$ (IO) & $8.7^{\circ}$ & $8.5^{\circ}$ - $8.9^{\circ}$ & $8.2^{\circ}$ - $9.1^{\circ}$ & $8.0^{\circ}$ - $9.4^{\circ}$ \\
  \hline
  $\delta$ (NO) & $254^{\circ}$ & $182^{\circ}$ - $353^{\circ}$& $0^{\circ}$ - $360^{\circ}$ & $0^{\circ}$ - $360^{\circ}$ \\
  \hline
  $\delta$ (IO) &$266^{\circ}$& $210^{\circ}$ - $322^{\circ}$ & $0^{\circ}$ - $16^{\circ}$ $\oplus$ $ 155^{\circ}$ - $360^{\circ}$  & $0^{\circ}$ - $360^{\circ}$ \\
\hline
\end{tabular}
\caption{\label{tab2}Current neutrino oscillation parameters from global fits at 1$\sigma$, 2$\sigma$ and 3$\sigma$ confidence level \cite{6}. NO (IO) refers to normal (inverted) neutrino mass ordering.}
\end{center}
\end{small}
\end{table}

\section{Numerical analysis}
The experimental constraints on neutrino parameters at 1$\sigma$, 2$\sigma$ and 3$\sigma$ confidence level (CL) are given in Table \ref{tab2}.
The effective Majorana mass term relevant for neutrinoless double beta ($0\nu\beta\beta$) decay  is given by
\begin{equation}
|M_{ee}|=|m_{1}c_{12}^{2}c_{13}^{2}e^{2i\rho}+m_{2}s_{12}^{2}c_{13}^{2}e^{2i\sigma}+m_{3}s_{13}^{2}|. 
\end{equation}
Observation of $0\nu\beta\beta$ decay will imply lepton number violation and Majorana nature of neutrinos. For reviews on $0\nu\beta\beta$ decay see Ref. \cite{33,34}. A large number of projects such as CUORICINO \cite{35}, CUORE \cite{36}, GERDA \cite{37}, MAJORANA \cite{38}, SuperNEMO \cite{39}, EXO \cite{40},GENIUS\cite{41} aim to achieve a sensitivity upto 0.01eV for $|M_{ee}|$. We take the upper limit on $|M_{ee}|$ to be 0.5 eV \cite{34}. Data from the Planck satellite \cite{42} combined with other cosmological data put a limit on the sum of neutrino masses as
\begin{equation}
\Sigma = \sum_{i = 1}^3 m_{i} < 0.23 \textrm{eV \ \ \ at 95\% CL.}
\end{equation}
In the present analysis, we assume a more conservative upper limit $\Sigma< 1$ eV, on the sum of neutrino masses.\\
Since we are considering one of the neutrino masses to be zero, we have two possibilities \textit{i.e.} either $m_{1}=0$ or $m_{3}=0$ corresponding to normal or inverted mass orderings, respectively. Eqs. (15) and (20) incorporate the constraints of a vanishing neutrino mass and one texture zero, for normal and inverted mass orderings, respectively. We span the parameter space of input neutrino oscillation parameters ($\theta_{12}$, $\theta_{23}$, $\theta_{13}$, $\Delta m^{2}$, $\Delta m^{2}$) lying in their $3\sigma$ ranges by randomly generating points of the order of $10^{7}$. Since the Dirac CP-violating phase $\delta$ is experimentally unconstrained at $3\sigma$ level, therefore, we vary $\delta$ within its full possible range [$0^\circ$, $360^{\circ}$]. Using Eqs. (15) and (20) and the experimental inputs on neutrino mixing angles and mass-squared differences, the parameter space of $\delta$, $\rho$, $\sigma$ and $|M_{ee}|$ can be subsequently constrained.\\
It is found from our analysis that for one texture zero neutrino mass matrices with $m_{1}=0$, all the six patterns are now inconsistent with the latest global fits of neutrino oscillation data at $3\sigma$ level. This result is mainly due to the reduced 3$\sigma$ errors on the experimentally measured values of $\theta_{13}$. On the other hand, for singular one texture zero neutrino mass matrices with $m_{3}=0$, only Classes $P_{2}$, $P_{3}$, $P_{4}$, $P_{5}$ are found to be compatible with the latest experimental data at $3\sigma$ level. We have also performed the numerical analysis by considering 1$\sigma$ and 2$\sigma$ ranges of neutrino mixing angles and mass-squared differences.
\begin{center}
\textbf{Case I: $m_{1}=0$ (normal mass ordering)}
\end{center}
Classes $P_{2},P_{3}, P_{6}$ with $m_{1}=0$  have already been ruled out in Ref. \cite{17}. In the following we check the viability of the rest of the classes viz. $P_{1},P_{4}, P_{5}$ with current experimental data.\\

\textbf{Class $P_{1}$:} Using Eq. (15), the exact analytical expression of $R_{\nu}$ in terms of $\theta_{13}$ is given by
\begin{equation}
R_{\nu} =\frac{{t}_{13}^{4}}{({s}_{12}^{4}-{t}_{13}^{4})},
\end{equation}
where $t_{13} = \tan \theta_{13}$. With the experimentally allowed ranges of $\theta_{12}$ and $\theta_{13}$, we find that $R_{\nu}$ turns out to be below its experimentally allowed 3$\sigma$ range and hence is in conflict with the latest data.\\

\textbf{Class $P_{4}$:} With the help of Eq. (15), we obtain the following expression for $R_\nu$ in the leading order approximation of $s_{13}$
\begin{equation}
R_{\nu} \approx \frac{{t}_{23}^{2}
s_{13}^{2}}{{s}_{12}^{2}{c}_{12}^{2}}.
\end{equation}
The latest mixing data leads to rather higher values of $R_{\nu}$, lying in the range (0.05 - 0.5) as compared to the allowed 3$\sigma$ range (0.0279 - 0.0372) of $R_\nu$ and hence class $P_4$ is excluded by current experimental data at 3$\sigma$ CL. \\

\textbf{Class $P_{5}$ :} We obtain the following expression for $R_\nu$ in the leading order approximation of $s_{13}$
\begin{equation}
R_{\nu} \approx \frac{
s_{13}^{2}}{{t}_{23}^{2} {s}_{12}^{2}{c}_{12}^{2}}.
\end{equation}
Since classes $P_{5}$ and $P_{4}$ are related via permutation symmetry given in Eq. (8), their phenomenological implications are similar. As in the case of class $P_4$ the latest mixing data leads to values of $R_{\nu}$ above its experimentally allowed 3$\sigma$ range hence, class $P_5$ is also ruled out by the latest experimental data. Ref. \cite{28} has also found classes $P_4$ and $P_5$ to be incompatible with the recent data for normal mass ordering. Further, in Ref. \cite{29} it has been shown that these classes remain disfavoured even after taking into account the renormalization group effects.\\

\begin{figure}[h!]
\begin{center}
\subfigure[]{\includegraphics[width=0.4\columnwidth]{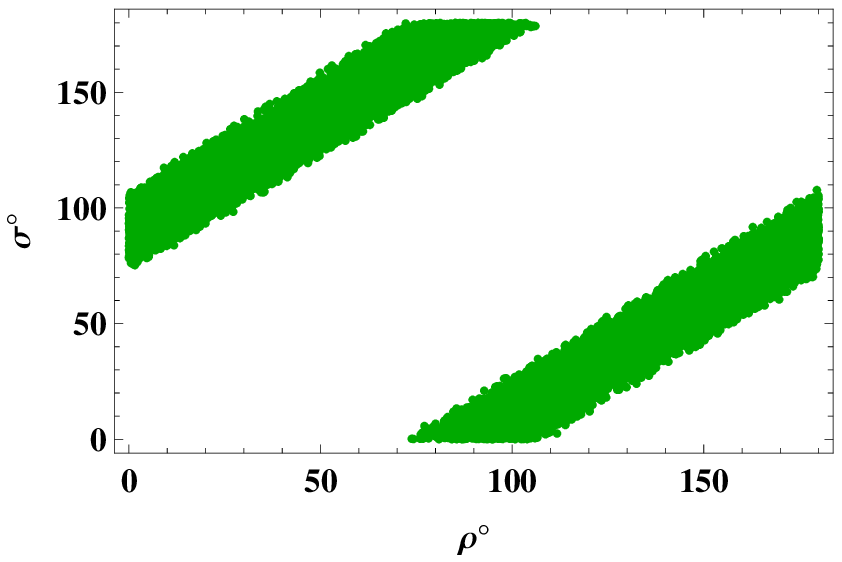}} \ \ \ 
\subfigure[]{\includegraphics[width=0.4\columnwidth]{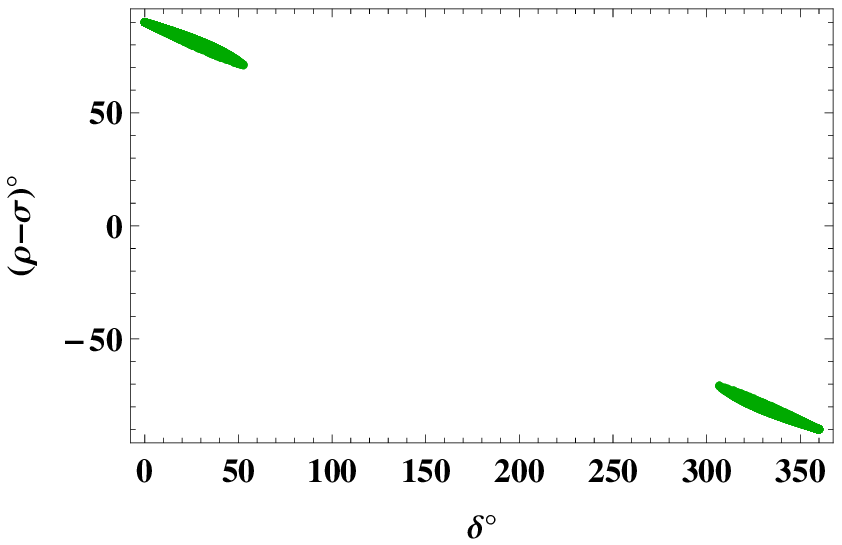}}\\
\subfigure[]{\includegraphics[width=0.4\columnwidth]{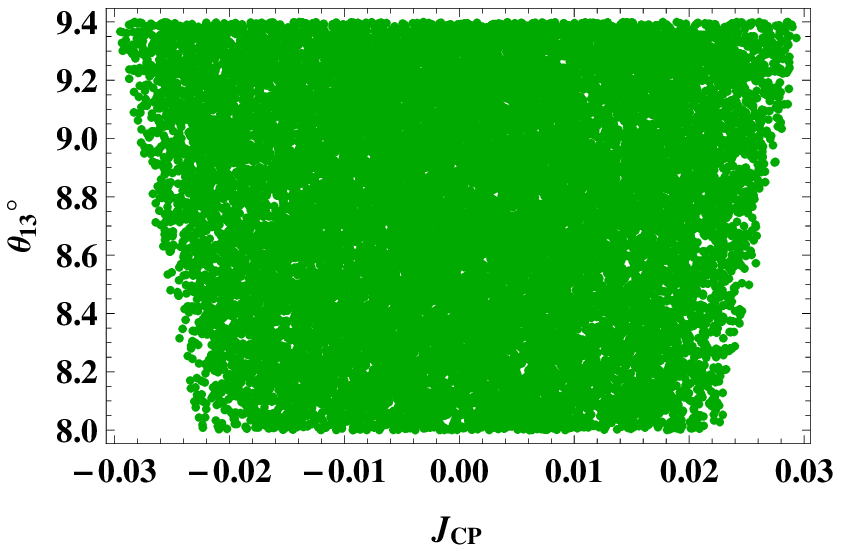}} \ \ \ 
\subfigure[]{\includegraphics[width=0.4\columnwidth]{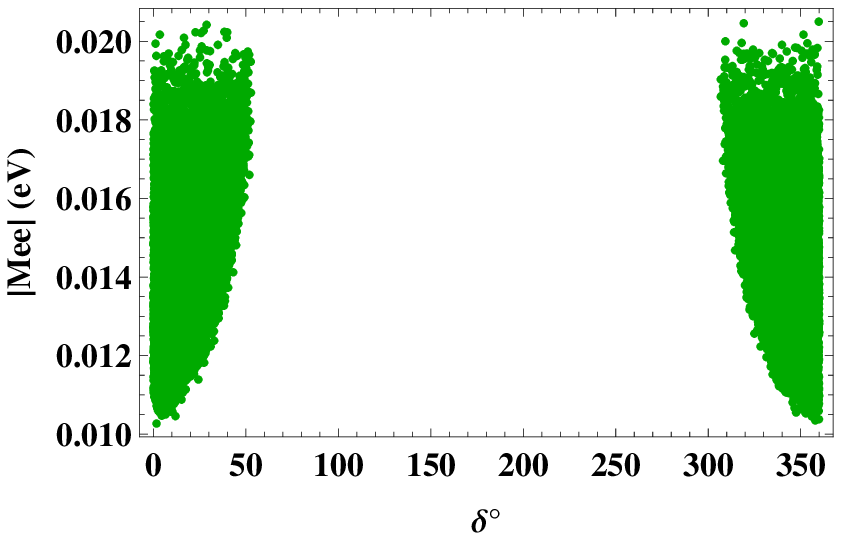}}
\caption{\label{fig1} Correlation plots for class $P_2$}
\end{center}
\end{figure}

\begin{figure}[h!]
\begin{center}
\subfigure[]{\includegraphics[width=0.4\columnwidth]{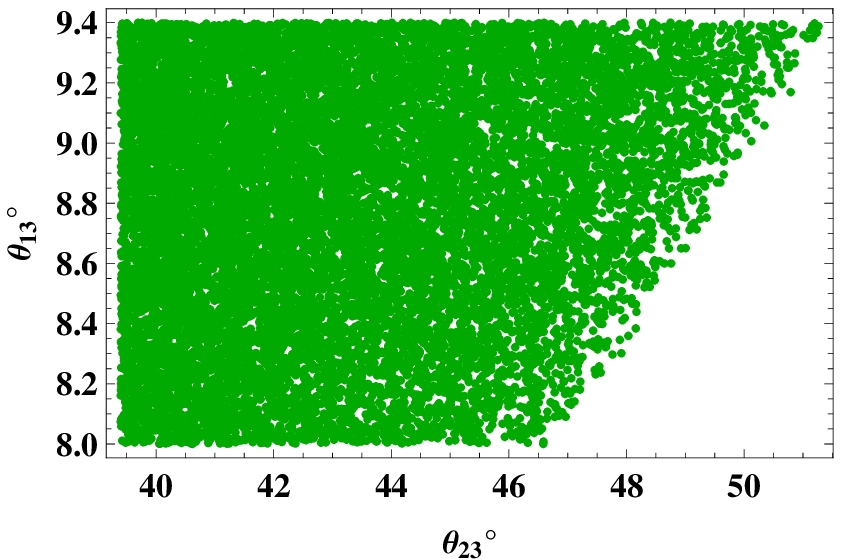}} \ \ \ 
\subfigure[]{\includegraphics[width=0.4\columnwidth]{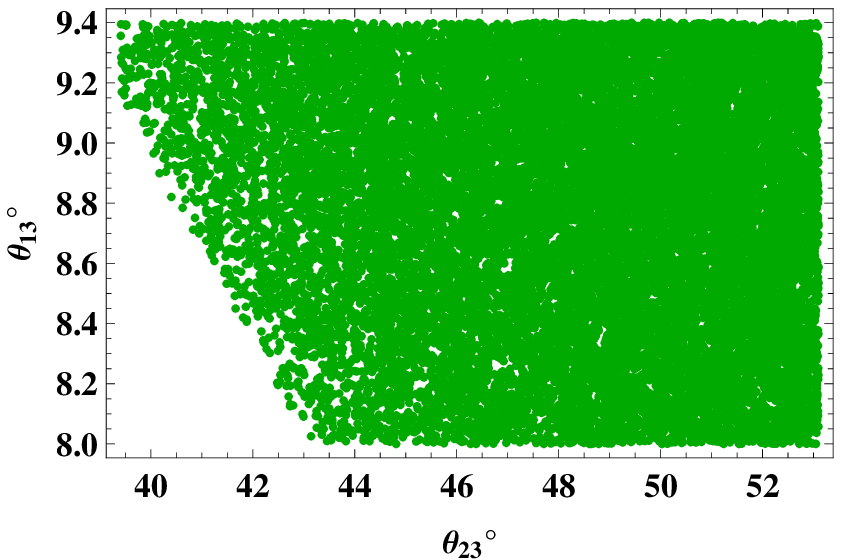}}\\
\caption{\label{fig2} Correlation plots for classes $P_{3}$ (a) and $P_{2}$ (b) depicting the 2-3 interchange symmetry. }
\end{center}
\end{figure}

\begin{figure}[h!]
\begin{center}
\subfigure[]{\includegraphics[width=0.4\columnwidth]{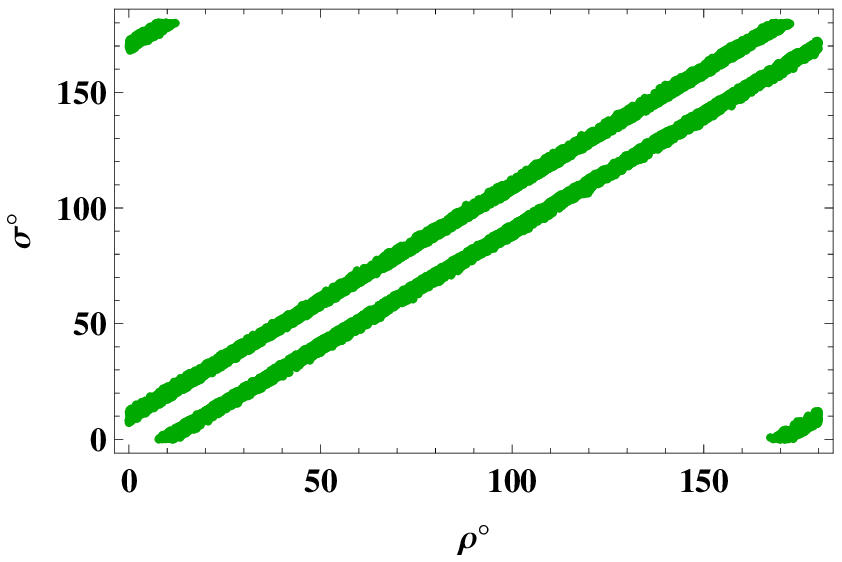}} \ \ \ 
\subfigure[]{\includegraphics[width=0.4\columnwidth]{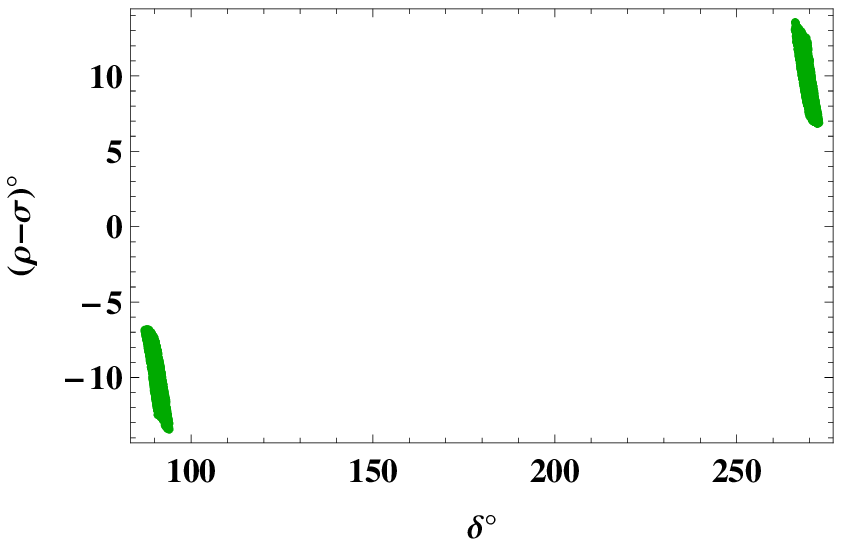}}\\
\subfigure[]{\includegraphics[width=0.4\columnwidth]{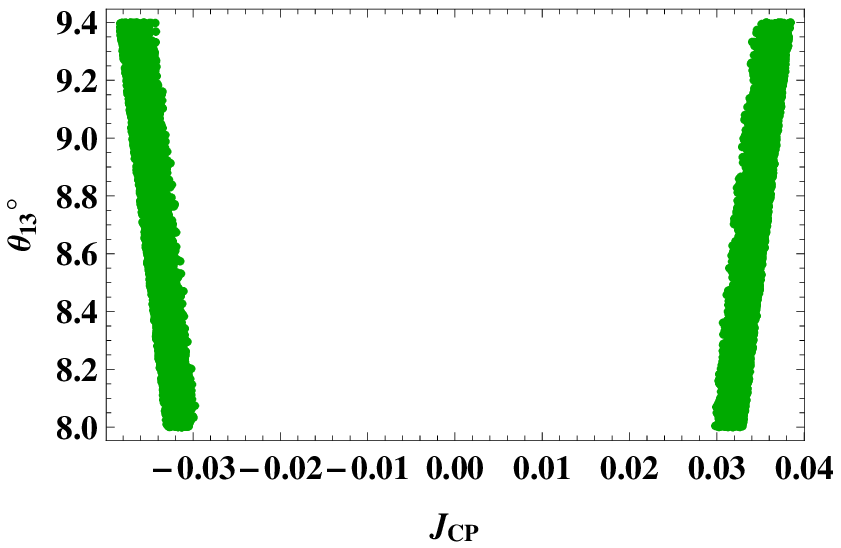}} \ \ \ 
\subfigure[]{\includegraphics[width=0.4\columnwidth]{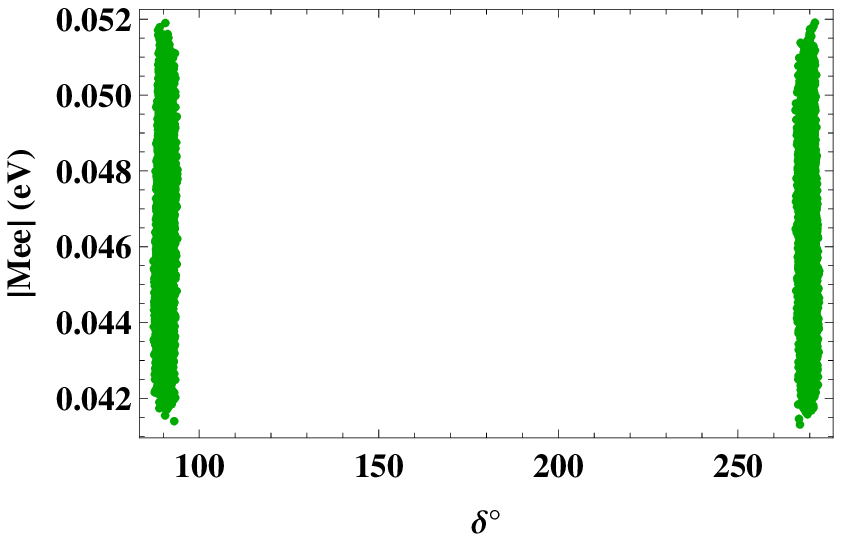}}
\caption{\label{fig3}Correlation plots for class $P_4$}
\end{center}
\end{figure}

Classes $P_4$ and $P_5$ with normal mass ordering have been studied in Ref. \cite{20} with the additional constraint of CSD2 \cite{20, 21}. In Ref. \cite{22}, classes $P_1$, $P_4$ and $P_5$ with normal mass ordering have been examined with the outcome that the (1,3) element of the neutrino mixing matrix and the parameter $R_\nu$ are found to be linked for these classes.\\
All these classes are now incompatible with the recent data \cite{6} for the normal mass ordering and one has to consider modifications to these classes for them to be compatible with the latest data. In this direction, small charged lepton contributions to classes $P_4$ and $P_5$ along with the CSD2 constraint have been considered in Ref. \cite{21} where, a unified indirect family symmetry model has been constructed, which leads to $\theta_{13} \sim 8^\circ$ - $9^\circ$. 
\begin{center}
\textbf{Case II:  $m_{3}=0$ (inverted mass ordering)}
\end{center}
For the inverted neutrino mass ordering case, classes $P_{2}$, $P_{3}$, $P_{4}$, $P_{5}$ are found to be compatible with the
latest neutrino oscillation data at $3\sigma$ CL.  Interestingly, classes $P_{2}$ and $P_{3}$ cannot satisfy the experimental data at $1\sigma$ CL while remaining classes $P_{4}$ and $P_{5}$ predict the Dirac phase $\delta$ to be near $270^{\circ}$ at the same confidence level. The correlation plots for these classes are given in figures \ref{fig1} - \ref{fig3} using 3$\sigma$ ranges of the known neutrino oscillation parameters while numerical results at 1$\sigma$, 2$\sigma$ and 3$\sigma$ CL are given in Table \ref{tab2}. Discussed below are the phenomenological implications of the experimentally allowed classes.\\
\textbf{Class $P_{2}$:} To the leading order in $s_{13}$ we get the following expression for the mass ratio $\left(\frac{m_2}{m_1}\right)$
\begin{equation}
\frac{m_{2}}{m_{1}}\approx t_{12}^{2}\bigg(1+\frac{2
c_{\delta}s_{13}t_{23}}{s_{12}c_{12}}\bigg),
\end{equation}
Some of the interesting plots for class $P_2$ are shown in Figs. \ref{fig1} and \ref{fig2}. In Fig. \ref{fig1}(a) we have shown the correlation plot between the two Majorana phases $\rho$ and $\sigma$. In the case of $m_3 = 0$, one has the freedom to make an overall phase rotation of neutrino masses so that the Majorana phase associated with one of the non-zero neutrino masses \textit{i.e.} $m_1$ or $m_2$ may be rotated away and we are left with only one phase difference which is physical. In Fig. \ref{fig1}(b) we have shown the correlation plot between the physical Majorana phase difference $(\rho - \sigma)$ and Dirac phase $\delta$. One can see that $(\rho - \sigma)$ and $\delta$ are constrained to small ranges for class $P_2$. A vanishing $\delta$ is still possible for class $P_2$ which allows for vanishing $J_{CP}$ along with non-zero values [Fig. \ref{fig1}(c)]. The effective Majorana mass $|M_{ee}|$ has been plotted against $\delta$ in Fig. \ref{fig1}(d). It is clear that only a narrow range for $|M_{ee}|$ $\sim$ (0.0102 - 0.0205) eV is allowed for class $P_2$.

\textbf{Class $P_{3}$:} Since class $P_3$ is related to class $P_2$ via permutation symmetry [Eq. (8)], the phenomenological implications for class $P_3$ can be obtained from class $P_2$ using Eq. (9). The 2-3 interchange symmetry between classes $P_2$ and $P_3$ is shown in figure \ref{fig2}. From Fig. \ref{fig2}(a) (\ref{fig2}(b)) one can see that for higher values of $\theta_{13}$, lower (upper) quadrant of $\theta_{23}$ is preferred for class $P_3$ $(P_2)$. Figs. \ref{fig2}(a) and (b) may appear to show slight deviation from the 2-3 interchange symmetry relation: 
\begin{equation}
\theta_{23}^{P_3}=90^{\circ}-\theta_{23}^{P_2}.
\end{equation}
However, this apparent deviation is just because the experimentally allowed 3$\sigma$ range for $\theta_{23}$ is not symmetric around $\theta_{23} = 45^\circ$.\\

\textbf{Class $P_{4}$:} We have the following expression for the mass ratio $\left(\frac{m_2}{m_1}\right)$ in leading order terms of $s_{13}$
\begin{equation}
\frac{m_{2}}{m_{1}}\approx \bigg( 1+ \frac{c_{\delta} s_{13}
t_{23}} {s_{12} c_{12}}\bigg).
\end{equation}
The correlation plots for class $P_4$ have been compiled in figure \ref{fig3}. Fig. \ref{fig3}(a) shows the correlation plot between the Majorana phases $\rho$ and $\sigma$. The physically relevant phase difference $(\rho - \sigma)$ has been plotted against Dirac phase $\delta$ in Fig. \ref{fig3}(b). Both $(\rho - \sigma)$ and $\delta$ have very small allowed ranges for class $P_4$. The allowed parameter space for $\delta$ is constrained near $90^\circ$ and $270^\circ$, which leads to the result that $J_{CP}$ cannot vanish for this class [Fig. \ref{fig3}(c)]. A very narrow range for $|M_{ee}|$ $\sim$ (0.0412 - 0.0520) eV is allowed for class $P_4$ [Fig. \ref{fig3}(d)].\\
\textbf{Class $P_{5}$:} As class $P_5$ is related to class $P_4$ via permutation symmetry [Eq. (8)], the phenomenological implications for class $P_5$ can be obtained from class $P_4$ using Eq. (9). The allowed values of $|M_{ee}|$ for classes $P_4$ and $P_5$ are on the larger side having no overlap with classes $P_2$ and $P_3$ [Table \ref{tab3}]. Thus, $|M_{ee}|$ can be used to distinguish between diagonal and off-diagonal one texture zero classes with a vanishing neutrino mass.
\begin{table}[h!]
\begin{footnotesize}
\noindent\makebox[\textwidth]{
\begin{tabular}{|c|c|c|c|c|c|c|}
\hline
Class& CL & $\delta$ & $\rho - \sigma$ & $|M_{ee}|$(eV)& $J_{CP}$ \\
\hline
$P_{1}$ & $1\sigma$ & $\times$& $\times$ & $\times$ & $\times$ \\
& $2\sigma$ & $\times$ & $\times$ & $\times$ & $\times$ \\
& $3\sigma$ & $\times$ & $\times$ & $\times$ & $\times$ \\
\hline
 $P_{2}$ & $1\sigma$ & $\times$ & $\times$ & $\times$ & $\times$ \\
 &$2\sigma$& $0^{\circ}$ - $16^{\circ}$ $\oplus$ $317^{\circ}$ - $360^{\circ}$ & $(-90.00^{\circ})$ - $(-74.80^{\circ})$ $\oplus$ $83.82^{\circ}$ - $90.00^{\circ}$& $0.0122$ - $0.0187$ & $(-0.0240)$ - $0.0102$\\
 & $3\sigma$& $0^{\circ}$ - $53^{\circ}$ $\oplus$ $306^{\circ}$ - $360^{\circ}$ & $(-90.0^{\circ})$ - $(-70.5^{\circ})$ $\oplus$ $71.0^{\circ}$ - $90.0^{\circ}$ & $0.0102$ - $0.0205$ &$(-0.03)$ - $0.03$\\  
\hline
 $P_{3}$ & $1\sigma$ & $\times$ & $\times$ & $\times$ & $\times$ \\
 & $2\sigma$ & $154.7^{\circ}$ - $214.0^{\circ}$ & $(-90.0^{\circ})$ - $(-81.5^{\circ})$ $\oplus$ $78.5^{\circ}$ - $90.0^{\circ}$ & $0.0123$ - $0.0167$ &$(-0.0206)$ - $0.0156$  \\
 & $3\sigma$ & $130^{\circ}$ - $230^{\circ}$ & $(-90^{\circ})$ - $(-73^{\circ})$ $\oplus$ $73^{\circ}$ - $90^{\circ}$ & $0.0102$ - $0.0187$ & $(-0.029)$ - $0.029$ \\
\hline
$P_{4}$ & $1\sigma$ & $267.8^{\circ}$ - $269.9^{\circ}$ &  $9.45^{\circ}$ - $11.55^{\circ}$ & $0.0446$ - $0.0482$ & $(-0.0356)$ - $(-0.0327)$ \\
& $2\sigma$ & $266.8^{\circ}$ - $271.7^{\circ}$ & $7.4^{\circ}$ - $12.5^{\circ}$ & $0.0428$ - $0.0501$ & $(-0.0369)$ - $(-0.0308)$ \\
& $3\sigma$ & $87.2^{\circ}$ - $94.2^{\circ}$ $\oplus$ $265.9^{\circ}$ - $272.9^{\circ}$ & $(-13.5^{\circ})$ - $(-6.7^{\circ})$ $\oplus$ $6.7^{\circ}$ - $13.5^{\circ}$ & $0.0412$ - $0.0520$ & $(-0.0385)$ - $(-0.0295)$ $\oplus$ $0.0295$ - $0.0385$\\
\hline
$P_{5}$ & $1\sigma$&  $268.5^{\circ}$ - $270.6^{\circ}$ &  $(-8.90^{\circ})$ - $(-7.26^{\circ})$&  $0.0449$ - $0.0485$ & $(-0.0560)$ - $(-0.0327)$ \\
& $2\sigma$ &  $267.6^{\circ}$ - $272.3^{\circ}$  &  $(-11.2^{\circ})$ - $(-6.6^{\circ})$ & $0.0431$ - $0.0502$ & $(-0.0369)$ - $(-0.0309)$ \\
& $3\sigma$ & $86.5^{\circ}$ - $93.5^{\circ}$ $\oplus$ $266.5^{\circ}$ - $273.5^{\circ}$  & $(-12.5^{\circ})$ - $(-6^{\circ})$ $\oplus$ $6^{\circ}$ - $12.5^{\circ}$ & $0.0413$ - $0.0521$ & $(-0.0385)$ - $(-0.0295)$ $\oplus$ $0.0295$ - $0.0385$\\
\hline
$P_{6}$ & $1\sigma$ & $\times$ & $\times$ & $\times$ & $\times$ \\
& $2\sigma$ & $\times$ & $\times$ & $\times$ & $\times$ \\
& $3\sigma$ & $\times$ & $\times$ & $\times$ & $\times$ \\
  \hline
\end{tabular}}
\caption{\label{tab3}The allowed ranges of Dirac CP-violating phase $\delta$, the physical Majorana phase difference $(\rho - \sigma)$, effective Majorana mass $|M_{ee}|$ and Jarlskog rephrasing invariant $J_{CP}$ for the experimentally allowed classes with $m_{3}=0$.}
\end{footnotesize}
\end{table} 
\section{Symmetry Realization}
Singular one texture zero neutrino mass matrices can be realized using a discrete Abelian flavor symmetry within the context of type-I seesaw mechanism \cite{26}. Such texture structures have been realized earlier in Ref.\cite{17} using $Z_{12} \times Z_2$ symmetry, requiring six (seven) $SU(2)_L$ doublet Higgses for the classes where the texture zero is present on the diagonal (off-diagonal) elements. Here, we show how the phenomenologically allowed singular one texture zero classes can be realized with a much smaller number of Higgs doublets and a smaller symmetry group. For the classes where the texture zero corresponds to diagonal elements we need only two Higgs doublets and for the off-diagonal texture zero classes only three Higgs doublets are required. The symmetry group used to realize these texture structures is $Z_8$.\\
Within the framework of type-I seesaw mechanism the effective neutrino mass matrix is given by
\begin{equation}
M_\nu \approx M_D M_R^{-1} M_D^T
\end{equation} 
where $M_D$ and $M_R$ are the Dirac and the right-handed neutrino mass matrices, respectively.
To realize the texture structures within the framework of type-I seesaw mechanism we extend the standard model by adding three right-handed neutrinos ($\nu_{eR}, \nu_{\mu R}, \nu_{\tau R}$) and a $SU(2)_L$ singlet scalar ($\chi$). \\
For Illustration, we show in detail the symmetry realization of class $P_2$ where the (2,2) element of $M_\nu$ is zero.
For class $P_2$ we assume the following transformation properties of the leptonic fields under the cyclic group $Z_8$
\begin{align}
 \overline{D}_{eL}&\rightarrow \omega^7 \overline{D}_{eL},& e_R & \rightarrow \omega e_R,& \nu_{e R}&\rightarrow \nu_{e R},&  \nonumber \\ \overline{D}_{\mu L}& \rightarrow \omega^4 \overline{D}_{\mu_L},& \mu_R & \rightarrow \omega^4 \mu_R ,&  \nu_{\mu R} & \rightarrow  \omega^4 \nu_{\mu R},&  \\ \overline{D}_{\tau L} & \rightarrow \omega^5 \overline{D}_{\tau L},& \tau_R & \rightarrow \omega^5 \tau_R ,&  \nu_{\tau R} & \rightarrow \omega \nu_{\tau R},& \nonumber
\end{align}
where $\omega$ = $e^{i 2 \pi/8}$, $D_{fL}$ $(f = e, \mu, \tau)$ denote $SU(2)_L$ doublets and $l_{fR}$, $\nu_{fR}$ denote the right-handed $SU(2)_L$ singlet charged lepton and neutrino fields, respectively. According to the above transformations of the leptonic fields, the bilinears $\overline{D}_{f L} l_{gR}$, $\overline{D}_{f L}\nu_{g R}$ and $\nu_{f R}^T C^{-1} \nu_{g R}$ relevant for $M_l$, $M_D$ and $M_R$, respectively, transform as
\begin{small} 
\begin{equation}
\overline{D}_{fL} l_{gR}  \sim \left(
\begin{array}{ccc}
1& \omega^3 &\omega^4 \\
\omega^5 &1 &\omega \\
\omega^6 & \omega &\omega^2
\end{array}
\right), \ \ \ \  \overline{D}_{fL} \nu_{g R} \sim \left(
\begin{array}{ccc}
\omega^7 & \omega^3 &1 \\
\omega^4 &1 &\omega^5 \\
\omega^5 & \omega & \omega^6
\end{array}
\right), \ \ \ \ \nu_{f R}^T C^{-1} \nu_{g R} \sim \left(
\begin{array}{ccc}
1& \omega^4 &\omega \\
\omega^4 & 1 &\omega^5 \\
\omega & \omega^5 & \omega^2
\end{array}
\right).
\end{equation}
\end{small}
We introduce two $SU(2)_L$ doublet Higgs ($\phi_1, \phi_2$) transforming as: $\phi_1 \rightarrow \phi_1$ and $\phi_2 \rightarrow \omega^6 \phi_2$ under the action of $Z_8$. These transformation properties of $\phi_1$ and $\phi_2$ will lead to the following $Z_8$ invariant Yukawa Lagrangian for class $P_2$
\begin{align} 
-\mathcal{L}_Y = \ & Y_1 (\overline{D}_{e L} e_R) \phi_1 + Y_2(\overline{D}_{\mu L} \mu_R) \phi_1 + Y_3(\overline{D}_{\tau L} \tau_R) \phi_2 + Y_4 (\overline{D}_{e L} \nu_{\tau R}) \tilde{\phi_1} \nonumber \\ \ & + Y_5 (\overline{D}_{\mu L} \nu_{\mu R}) \tilde{\phi_1} + Y_6 (\overline{D}_{\tau L} \nu_{\tau R}) \tilde{\phi_2} + \ \textrm{H. c.}
\end{align}
where $\tilde{\phi_j} = i \tau_2 \phi_j^*$ ($j = 1,2$).
When the Higgs fields ($\phi_j$) acquire non-zero vacuum expectation values $\langle \phi_j \rangle_o \neq 0$, we get the charged lepton mass matrix $M_l$ and the Dirac neutrino mass matrix $M_D$ of the following form
 \begin{align}
 M_l &= \left(
\begin{array}{ccc}
m_e & 0 & 0 \\  0 & m_\mu & 0 \\ 0 & 0 & m_\tau
\end{array}
\right),\\
 M_D &= \left(
\begin{array}{ccc}
0 & 0 & a \\  0 & b & 0 \\ 0 & 0 & c
\end{array}
\right),
 \end{align}
where $m_e = Y_1 \langle \phi_1 \rangle_o$, $m_\mu = Y_2 \langle \phi_1 \rangle_o $, $m_\tau = Y_3 \langle \phi_2 \rangle_o$, $a = Y_4 \langle \phi_1^* \rangle_o$, $b = Y_5 \langle \phi_1^* \rangle_o$ and $c = Y_6 \langle \phi_2^* \rangle_o$.
For the right-handed Majorana neutrino mass matrix, we assume a $SU(2)_L$ singlet scalar ($\chi$) transforming as $\chi \rightarrow \omega^3 \chi$ under $Z_8$, thus $\chi$ will lead to non-zero (2,3) element of $M_R$. Also, non-zero (1,1) and (2,2) elements of $M_R$ arise from bare Majorana mass terms which are already invariant under $Z_8$. This leads to the following form of $M_R$
\begin{equation}
 M_R = \left(
\begin{array}{ccc}
A & 0 & 0 \\  0 & B & C \\ 0 & C & 0
\end{array}
\right).
\end{equation}
Using the type-I seesaw mechanism, these $M_D$ and $M_R$ lead to an effective neutrino mass matrix having a zero (2,2) element and a vanishing neutrino mass (class $P_2$).\\
By assigning suitable transformation properties to the leptonic and Higgs fields under the action of $Z_8$, one can realize the remaining phenomenologically allowed classes of singular neutrino mass matrices with one texture zero. The structures of $M_D$ and $M_R$ for all the phenomenologically allowed classes are summarized in Table \ref{tab4}. The leptonic and Higgs field transformation properties under $Z_8$, leading to all the viable one texture zero classes with a vanishing neutrino mass are given in Table \ref{tab5}.  
\begin{table}[t!]
\begin{small}
\noindent\makebox[\textwidth]{
\begin{tabular}{|c|c|c|}
\hline Class & $M_D$ & $M_R$ \\
\hline $P_2$ & $\left(
\begin{array}{ccc}
0 & 0 & a \\  0 & b & 0 \\ 0 & 0 & c
\end{array}
\right)$ & $\left(
\begin{array}{ccc}
A & 0 & 0 \\  0 & B & C \\ 0 & C & 0
\end{array}
\right)$ \\
\hline $P_3$ & $\left(
\begin{array}{ccc}
a & 0 & 0 \\  b & 0 & 0 \\ 0 & 0 & c
\end{array}
\right)$ & $\left(
\begin{array}{ccc}
0 & 0 & A \\  0 & B & 0 \\ A & 0 & C
\end{array}
\right)$ \\
\hline $P_4$ & $\left(
\begin{array}{ccc}
a & 0 & 0 \\  0 & 0 & b \\ c & 0 & d
\end{array}
\right)$ & $\left(
\begin{array}{ccc}
A & 0 & 0 \\  0 & B & 0 \\ 0 & 0 & C
\end{array}
\right)$  \\
\hline $P_5$ & $\left(
\begin{array}{ccc}
a & 0 & 0 \\  b & 0 & c \\ 0 & 0 & d
\end{array}
\right)$ & $\left(
\begin{array}{ccc}
A & 0 & 0 \\  0 & B & 0 \\ 0 & 0 & C
\end{array}
\right)$ \\
\hline
\end{tabular}}
\end{small}
\caption{\label{tab4}Structures of $M_{D}$ and $M_{R}$ leading to experimentally allowed singular one texture zero effective neutrino mass matrices.}
\end{table}

\begin{table}[t!]
\begin{small}
\noindent\makebox[\textwidth]{
\begin{tabular}{|c|c|c|c|c|c|c|c|}
\hline Class & $\overline{D}_{eL}$, \ $\overline{D}_{\mu L}$, \ $\overline{D}_{\tau L}$ & $e_R$, \ $\mu_R$, \ $\tau_R$ & $\nu_{e R}$, \ $\nu_{\mu R}$, \ $\nu_{\tau R}$ & $\phi_1$ &$\phi_2$ & $\phi_3$ &  $\chi$ \\
\hline $P_2$ & $\omega^7$, \ $\omega^4$, \ $\omega^5$ & $\omega$, \ $\omega^4$, \ $\omega^5$  &  $1$, \ $\omega^4$, \ $\omega$  & $1$ &$\omega^6$ & $-$  &$\omega^3$ \\
\hline $P_3$ & $\omega^7$, \ $\omega^5$, \ $\omega^4$   & $\omega$, \ $\omega^5$, \ $\omega^4$ &  $\omega$, \ $1$, \ $\omega^4$ & $1$ & $\omega^6$ & $-$  &$\omega^3$ \\
\hline $P_4$ & $1$, \ $\omega^6$, \ $\omega^3$ & $1$, \ $\omega$, \ $\omega^4$ & $1$, \ $\omega^4$, \ $\omega^5$  & $1$ & $\omega$ & $\omega^3$  & $\omega^6$\\
\hline $P_5$ & $1$, \ $\omega^3$, \ $\omega^6$ & $1$, \ $\omega^4$, \ $\omega$ & $1$, \ $\omega^4$, \ $\omega^5$  & $1$ & $\omega$ & $\omega^3$  & $\omega^6$\\
\hline
\end{tabular}}
\end{small}
\caption{\label{tab5} Transformation properties of lepton and scalar fields under $Z_8$ for classes $P_2$, $P_3$, $P_4$ and $P_5$.}
\end{table}
\section{Summary}
We have done a systematic analysis of all the one texture zero neutrino mass matrices with a vanishing neutrino mass using the latest global fits of neutrino oscillation parameters. We find that all the six classes with normal mass ordering are now ruled out at $3\sigma$ confidence level whereas in case of inverted mass ordering only four classes ($P_{2}, P_{3}, P_{4}, P_{5}$) out of total six are consistent with the latest experimental data. Furthermore, only classes $P_4$ and $P_5$ are found to be compatible with the latest data at 1$\sigma$ confidence level. For classes $P_{4}$ and $P_{5}$ with inverted mass ordering, the parameter space of Dirac CP-violating phase $\delta$ is restricted to values near $\delta \approx 90^{\circ}$, $270^{\circ}$ ($\delta \approx 270^{0}$) at $3\sigma$ ($ 1\sigma$) confidence level. We have shown how the experimentally allowed classes can be realized within the context of type-I seesaw mechanism using $Z_8$ discrete symmetry and a small number of Higgs doublets. Classes where the texture zero corresponds to diagonal elements can be distinguished from the off-diagonal one texture zero classes on the basis of allowed 3$\sigma$ ranges of effective Majorana mass for these classes. For all the experimentally allowed classes we get ranges for the effective Majorana mass, which lie within the sensitivity limits of future neutrinoless double beta decay experiments.\\

\acknowledgements{
The research work of R. R. G. is supported by the Department of Science and Technology, Government of India, under Grant No. SB/FTP/PS-128/2013. M. G. would like thank CSIR for the support through research grant no 03(1313)/14/EMR-II.}

\end{document}